\documentclass[a4paper,12pt]{article}
\usepackage{epsfig}
\usepackage[margin=2cm]{geometry}
\usepackage{graphicx}
\usepackage{dcolumn}
\usepackage{bm}
\usepackage{longtable}
\usepackage{amssymb}
\usepackage{amsfonts}
\usepackage{amsmath}
\usepackage{yfonts}
\usepackage{slashed}

\allowdisplaybreaks[1]

\newcommand{\be}{\begin{equation}}
\newcommand{\ee}{\end{equation}}
\newcommand{\ba}{\begin{align}}
\newcommand{\ea}{\end{align}}
\newcommand{\one}{{\rm 1\kern -.9mm l}}



\begin{document}

\begin{titlepage}
\begin{flushright}
DFPD-13/TH/16\\
\par\end{flushright}
\vskip 2cm
\begin{center}
\textbf{\huge \bf Fermions, Wigs, and Attractors}
\\\vspace{.5cm}
\textbf{\vspace{2cm}}\\
{\Large L.G.C.~Gentile$ ^{~a, c, e,}$\footnote{lgentile@pd.infn.it},$\ $ P.A.~Grassi$ ^{~a, d,}$\footnote{pgrassi@mfn.unipmn.it}  A. Marrani$ ^{~f,}$\footnote{alessio.marrani@fys.kuleuven.be} and A.~Mezzalira$ ^{~b,}$\footnote{andrea.mezzalira@ulb.ac.be}}

\begin{center}
{a) { \it DISIT, Universit\`{a} del Piemonte Orientale,
}}\\
{{ \it via T. Michel, 11, Alessandria, 15120, Italy, }}
 \\ \vspace{.1cm}
 {b) { \it Physique Th\'eorique et Math\'ematique}}\\
{{ \it Universit\'e Libre de Bruxelles, C.P. 231, 1050 Bruxelles, Belgium,}}
 \\ \vspace{.1cm}
{c) { \it Dipartimento di Fisica Galileo Galilei,\\
Universit\`a di Padova,\\
via Marzolo 8, 35131 Padova, Italy,
}}
 \\ \vspace{.1cm}
 {d) { \it INFN - Gruppo Collegato di Alessandria - Sezione di Torino,}}
 \\ \vspace{.2cm}
 {e) { \it INFN, Sezione di Padova,\\
via Marzolo 8, 35131, Padova, Italy,}}
 \\ \vspace{.1cm}
{f) { \it ITF KU Leuven,\\
Celestijnenlaan 200D, 3001, Leuven, Belgium.}}
 \\ \vspace{.1cm}
\end{center}

\par\end{center}
\vfill{}

\begin{abstract}
{\vspace{.3cm}

       \noindent
We compute the modifications to the attractor mechanism due to fermionic corrections.
In ${\cal N}=2, D=4$ supergravity,  at the fourth order,
we find terms giving rise to new contributions to the horizon values of the scalar fields of the vector multiplets.
}
\end{abstract}
\vfill{}
\vspace{1.5cm}
\end{titlepage}

\vfill
\eject

\newpage \setcounter{footnote}{0}
\tableofcontents

\section{\label{Intro}Introduction}

The remarkable Schwarzschild solution to Einstein equations is the first
example of exact solution in general relativity. Since then, several
interesting solutions have been constructed with different properties, and a
number of theorems for black hole geometries has been proved. The search for
new solutions lived a new \textit{Renaissance} with the discovery of
supergravity: within this theory, Einstein equations are just a sector of a
broader framework, containing fermions and new matter fields. The latter are
sources of the gravitational field, but they are not generic since their
interactions are controlled by \textit{supersymmetry}. Consequently, for
such matter-gravity systems, new (BPS) solutions can be constructed, since
second-order partial-differential Einstein equations are replaced by
first-order ones, thus easier to solve. In that context, the solution to
supergravity equations of motion is generically constructed \textit{by
setting to zero all fermions}, while the bosonic fields acquire
non-vanishing v.e.v.'s.

For extremal black hole solutions, the \textit{attractor mechanism} \cite%
{AM-Refs} has been discovered; essentially, it states that the solution
computed at the horizon depends only upon the conserved charges of the
system, and it is independent of the value of the matter fields at infinity.
This is related to the \textit{no-hair theorem}, under which, for example, a
BPS black hole solution depends only upon its mass, its angular momentum and
other conserved charges. At the dawn of these studies, some Authors \cite%
{KL-1} posed the question whether the attractor mechanism has to be modified
in the presence of fermions. Their conclusion was that, at the level of
approximation of their computations, in the case of double-extremal BPS
solutions, the mechanism is unchanged. At the same time, \cite%
{Burrington:2004hf} investigated a similar issue for $\mathcal{N}=2, D=5$
AdS-black holes, and they found that the solution, as well as its asymptotic
charges, is modified at the first order due to fermionic contributions (even
though they did not study the attractor mechanism nor its possible
modifications).

All these studies followed the seminal paper by Aichelburg and Embacher \cite%
{Aichelburg:1986wv}, in which they started from a $\mathcal{N}=2, D=4$
asymptotically flat black hole solution and computed iteratively the
supersymmetric variations of the background in terms of the flat-space
Killing spinors. Due to the Grassmannian nature of the fermions, this
procedure ends up after a finite number of iterations, and the complete
solution can be constructed. In terms of the latter, the modifications to
the asymptotic charges were computed in \cite{Aichelburg:1986wv}. However,
once again, the attractor mechanism was not investigated.

Recently, some of the authors of the present investigation addressed the
same question starting from a different perspective, namely the AdS/CFT
correspondence between AdS black holes and strongly-interacting fluids on
the AdS boundary. They provided the complete fermionic solution (\textit{wig}%
) to non-extremal black holes in several dimensions \cite{Gentile:2011jt}.

Here, we present a complete computation of the fermionic corrections to
static, spherically symmetric, asymptotically flat, dyonic, BPS
double-extremal black holes of $\mathcal{N}=2,D=4$ supergravity. Differently
from \cite{KL-1}, we find that the scalar fields acquire a non-trivial
contribution at the fourth order of the fermionic expansion, leading to a
non-trivial modification of the attractor mechanism.

We would like to point out that we compute the wigging by performing a
perturbation of the unwigged purely bosonic (double) extremal BPS extremal
black hole solution; thus, within this approximation, we consider quantities
like the radius of the event horizon unchanged. The complete analysis,
including the study of the fully-backreacted wigged black hole metric, will
be presented elsewhere \cite{to-appear-1}.\bigskip

The plan of the paper is as follows.

In Sec.~\ref{mc} we introduce the simplest class of models of $\mathcal{N}=2$%
, $D=4$ Einstein ungauged supergravity coupled to Abelian vector multiplets,
namely the so-called \textit{minimally coupled} class.

The wigging correction of all fields in the gravity and vector multiplets is
then computed in Sec.~\ref{Wigging}, and the modification of the attractor
mechanism at the event horizon of the BPS double-extremal black hole
solution is derived in Sec.~\ref{AM-modified}.

Within the aforementioned approximation (\textit{i.e.}, disregarding the
backreaction), the simplest example, namely the axion-dilaton model and its
wigging, is studied in some detail in Sec.~\ref{Axion-Dilaton}.

The final Sec.~\ref{Conclusion} gives an outlook and mentions various
further future developments.


\section{\label{mc}\textit{Minimally Coupled} Maxwell-Einstein \newline
$\mathcal{N}=2$ Supergravity}

Namely, we consider $n$ Abelian vector multiplets \textit{minimally coupled}
to the $\mathcal{N}=2$, $D=4$ gravity multiplet \cite{Luciani}, in absence
of gauging and hypermultiplets. The complex scalar fields from the vector
multiplets coordinatize a class of symmetric special K\"{a}hler manifolds,
namely the non-compact complex projective spaces $\overline{\mathbb{CP}}^{n}$%
, characterized by the vanishing of the so-called $C$-tensor $C_{ijk}$ of
special K\"{a}hler geometry (cfr. \textit{e.g.} \cite{N=2-Big}, as well as
\cite{K-rev}, and Refs. therein). In turn, this implies the Riemann tensor
to enjoy the following expression in terms of the metric of the non-linear
sigma model ($i=1,...,n$):%
\begin{equation}
C_{ijk}=0\Rightarrow R_{i\overline{j}k\overline{l}}=-g_{i\overline{j}}g_{k%
\overline{l}}-g_{i\overline{l}}g_{k\overline{j}}.  \label{vanishing}
\end{equation}%
At least among the cases with symmetric scalar manifolds, minimally coupled
models are the only ones that admit \textquotedblleft
pure\textquotedblright\ supergravity by simply setting $n=0$.

By virtue of (\ref{vanishing}), minimally coupled models exhibit simple
properties, allowing for an explicit study of various solutions to the
equations of motion\footnote{%
For a treatment of the attractor mechanism \cite{AM-Refs} and marginal
stability in extremal black hole solutions of these models, see \textit{e.g.}
\cite{early-axion-dilaton, Gnecchi-1, FMO-MS-1}, and Refs. therein. For an
analysis of the duality orbits and related \textit{moduli spaces}, \textit{%
cfr.} \cite{BFGM-1,Small-Orbits,FM-2}. These models have also been treated
in \cite{ADFT-N=1}, and more recently in \cite{Ortin-1}.}.

This class of models can be seen as describing a \textit{multi-dilaton system%
} \cite{Hayakawa-1}; note, however, that they cannot be uplifted to $D=5$
(see \textit{e.g.} \cite{Gnecchi-1}), nor they can be obtained by standard
Calabi-Yau compactifications.

The case of only one vector multiplet ($n=1$) corresponds indeed to the
so-called \textit{axion-dilaton} system of $\mathcal{N}=2$ supergravity.
This will be treated in some detail in Sec.~\ref{Axion-Dilaton}.

Within this class, remarkable simplifications take place in the
supersymmetry transformations, which are reported below; the treatment of
more general models will be presented elsewhere \cite{to-appear-1}.

\section{\label{Wigging}The Wigging}

As mentioned above, we consider $\mathcal{N}=2,D=4$ Poincar\'{e}
supergravity minimally coupled to $n$ Abelian vector multiplets; as notation
and conventions, we adopt the ones of \cite{N=2-Big}. The supersymmetry
transformations for fermionic fields are
\begin{align}
\delta \psi _{A\,\mu }=& \nabla _{\mu }\epsilon _{A}-\frac{1}{4} \left(
\partial _{i}K\bar{\lambda}^{i\,B} \epsilon_{B} - \bar{\partial}_{\bar{\imath%
}}K\bar{\lambda}^{\bar{\imath}}_{B} \epsilon^{B} \right) \psi _{A\,\mu }+
\notag \\
& +\left( A_{A}{}^{\nu \,B}g_{\mu \nu }+A_{A}^{^{\prime }}{}^{\nu \,B}\gamma
_{\mu \nu }\right) \epsilon _{B}+  \notag \\
& +\varepsilon _{AB}T_{\mu \nu }^{-}\gamma ^{\nu }\epsilon ^{B} \ ,  \notag
\\
\delta \lambda ^{i\,A} =& \frac{1}{4}\left( \partial_{j}K \bar{\lambda}%
^{j\,B} \epsilon_{B} - \bar{\partial}_{\bar{j}}K\bar{\lambda}^{\bar{j}}_{B}
\epsilon^{B} \right) \lambda ^{i\,A}+  \notag \\
& -\Gamma ^{i}{}_{jk}\bar{\lambda}^{k\,B}\epsilon _{B}\lambda
^{j\,A}+i\left( \partial _{\mu }z^{i}-\bar{\lambda}^{i\,B}\psi _{B\mu
}\right) \gamma ^{\mu }\epsilon ^{A}+  \notag \\
& +G_{\mu \nu }^{i\,-}\gamma ^{\mu \nu }\epsilon _{B}\varepsilon
^{AB}+D^{i\,AB}\epsilon _{B}\ ,  \label{fermionicSUSY}
\end{align}%
while bosonic fields transform as
\begin{align}
\delta e_{\mu }^{a}=& -i\bar{\psi}_{A\,\mu }\gamma ^{a}\epsilon ^{A}-i\bar{%
\psi}^{A}{}_{\mu }\gamma ^{a}\epsilon _{A}\ ,  \notag \\
\delta A_{\mu }^{\Lambda }=& 2\bar{L}^{\Lambda }\bar{\psi}_{A\,\mu }\epsilon
_{B}\varepsilon ^{AB}+2L^{\Lambda }\bar{\psi}^{A}{}_{\mu }\epsilon
^{B}\varepsilon _{AB}+  \notag \\
& +i\left( f_{i}^{\Lambda }\bar{\lambda}^{i\,A}\gamma _{\mu }\epsilon
^{B}\varepsilon _{AB}+\bar{f}_{\bar{\imath}}^{\Lambda }\bar{\lambda}_{A}^{%
\bar{\imath}}\gamma _{\mu }\epsilon _{B}\varepsilon ^{AB}\right) \ ,  \notag
\\
\delta z^{i}=& \bar{\lambda}^{i\,A}\epsilon _{A}\,,  \label{bosonicSUSY}
\end{align}%
where the auxiliary fields $A_{A}{}^{\mu \,B}$, $A_{A}^{^{\prime }}{}^{\mu
\,B}$ are defined as
\begin{align}
A^{\mu }{}_{A}^{B}:=& -\frac{i}{4}g_{\bar{k}l}\left( \bar{\lambda}_{A}^{\bar{%
k}}\gamma ^{\mu }\lambda ^{l\,B}-\delta _{A}^{B}\bar{\lambda}_{C}^{\bar{k}%
}\gamma ^{\mu }\lambda ^{l\,C}\right) \ ,  \notag \\
{A^{^{\prime }}}{}^{\mu }{}_{A}^{B}:=& \frac{i}{4}g_{\bar{k}l}\left( \bar{%
\lambda}_{A}^{\bar{k}}\gamma ^{\mu }\lambda ^{l\,B}-\frac{1}{2}\delta
_{A}^{B}\bar{\lambda}_{C}^{\bar{k}}\gamma ^{\mu }\lambda ^{lC}\right) \ ,
\label{auxiliaryA}
\end{align}%
and the supercovariant field strength as
\begin{equation}
\widetilde{F}_{\mu \nu }^{\Lambda }:=\mathcal{F}_{\mu \nu }^{\Lambda
}+L^{\Lambda }\bar{\psi}_{\mu }^{A}\psi _{\nu }^{B}\varepsilon
_{AB}-if_{i}^{\Lambda }\bar{\lambda}^{i\,A}\gamma _{\left[ \nu \right. }\psi
_{\left. \mu \right] }^{B}\varepsilon _{AB}+\mathrm{h.c.}\,.
\end{equation}%
From the \text{Vielbein} postulate, the $\mathcal{N}=2$ spin connection
reads (\textit{cfr. e.g.} \cite{Samtl-lects})
\begin{equation}
{\omega _{\mu }^{ab}}=\frac{1}{2}{e_{c\mu }}\left[ \Omega ^{abc}-\Omega
^{bca}-\Omega ^{cab}\right] +K_{\phantom{a}\mu }^{a\phantom{\mu}b}\,,
\end{equation}%
where 
${\Omega ^{abc}:}=e^{\mu a}e^{\nu b}\left( \partial _{\mu }{e_{\nu }^{c}}%
-\partial _{\nu }{e_{\mu }^{c}}\right) 
$ 
and 
$K_{\phantom{a}\mu }^{a\phantom{\mu}b}:=-i\bar{\psi}_{A}^{\left[ a\right.
}\gamma ^{\left. b\right] }\psi _{\mu }^{A}-i\bar{\psi}^{Aa}\gamma ^{b}\psi
_{A\mu }$. 
For $\overline{\mathbb{CP}}^{n}$ models, various quantities of special
geometry \cite{N=2-Big} get simplified as follows:
\begin{align}
T_{\mu \nu }^{-}:=& 2i\left( \text{Im}{\mathcal{N}}\right) _{\Lambda \Sigma
}L^{\Sigma }\widetilde{F}_{\mu \nu }^{\Lambda -}\ ,  \notag \\
T_{\mu \nu }^{+}:=& 2i\left( \text{Im}{\mathcal{N}}\right) _{\Lambda \Sigma }%
\bar{L}^{\Sigma }\widetilde{F}_{\mu \nu }^{\Lambda +}\ ,  \notag \\
G_{\mu \nu }^{i-}:=& -g^{i\bar{j}}\bar{f}_{\bar{j}}^{\Gamma }\left( \text{Im}%
{\mathcal{N}}\right) _{\Gamma \Lambda }\widetilde{F}_{\mu \nu }^{\Lambda -}\
,  \notag \\
G_{\mu \nu }^{\bar{\imath}+}:=& -g^{\bar{\imath}j}f_{j}^{\Gamma }\left(
\text{Im}{\mathcal{N}}\right) _{\Gamma \Lambda }\widetilde{F}_{\mu \nu
}^{\Lambda +}\ ,  \notag \\
\mathcal{F}_{\mu \nu }^{\Lambda }:=& \partial _{\left[ \mu \right.
}A_{\left. \nu \right] }^{\Lambda }\ ,  \notag \\
\nabla \epsilon _{A}:=& \text{d}\epsilon _{A}-\frac{1}{4}\gamma _{ab}\omega
^{ab}\wedge \epsilon _{A}+\frac{i}{2}Q\wedge \epsilon _{A}\ ,  \notag \\
Q_{\mu }:=& -\frac{i}{2}\left( \partial _{i}K\partial _{\mu }z^{i}-\bar{%
\partial}_{\bar{\imath}}K\partial _{\mu }\bar{z}^{\bar{\imath}}\right) \ ,
\notag \\
D^{i\,AB}=& 0\,,  \label{otherQuantities}
\end{align}%
where $\omega ^{ab}$ is the spacetime spin connection, $Q$ is the connection
of the $U\left( 1\right) _{R}-$line bundle, $\omega _{A}^{\phantom{A}B}:=%
\frac{i}{2}\omega ^{x}\left( \sigma _{x}\right) _{A}^{\phantom{A}B}$ where $%
\omega ^{x}$ is the connection of the (global, in this case) $SU\left(
2\right) _{R}-$bundle and $\sigma _{x}$ are the $SU\left( 2\right) $ Pauli
matrices. Note also that $
\omega _{\phantom{A}B}^{A}:=\varepsilon ^{AC}\varepsilon _{DB}\omega _{C}^{%
\phantom{C}D}\,.$ 
Furthermore, the (anti)self-dual supercovariant field strength is defined as
\begin{equation}
\mathcal{F}_{\mu \nu }^{\Lambda \pm }:=\frac{1}{2}\left( \mathcal{F}_{\mu
\nu }^{\Lambda } \pm \frac{i}{2}\varepsilon _{\mu \nu \rho \sigma }\mathcal{F%
}^{\rho \sigma |\Lambda }\right) \,,
\end{equation}%
and the same holds for $\tilde{F}_{\mu \nu }^{\Lambda \pm }$. Note that $g$
is the determinant of the spacetime metric.

The following identities of the special geometry of $\overline{\mathbb{CP}}%
^{n}$ are used throughout:
\begin{align}
& f_{i}^{\Lambda }=\nabla _{i}L^{\Lambda }:=\left( \partial _{i}+\frac{1}{2}%
\partial _{i}K\right) L^{\Lambda }\ ,  \notag  \label{SPGE} \\
& \quad L^{\Lambda }=e^{\frac{K}{2}}X^{\Lambda }\ ,\quad \nabla
_{i}f_{j}^{\Lambda }=0\ ,  \notag \\
& \quad \nabla _{i}\bar{f}_{\bar{j}}^{\Lambda }=g_{i\bar{j}}\bar{L}^{\Lambda
}\ ,\quad \bar{\nabla}_{\bar{\imath}}L^{\Lambda }=0\ ,  \notag \\
& \mathrm{Im}~\mathcal{N}_{\Lambda \Gamma }f_{i}^{\Lambda }L^{\Gamma }=%
\mathrm{Im}~\mathcal{N}_{\Lambda \Gamma }\bar{f}_{\bar{\imath}}^{\Lambda }%
\bar{L}^{\Gamma }=0\,.
\end{align}%
\medskip

We now start with a purely bosonic background: the \textit{double-extremal} (%
$1/2$-)BPS black hole. For this solution, the near-horizon conditions \cite%
{AM-Refs}%
\begin{align}
&\partial _{\mu }z^{i}=0 \ , & G_{\mu \nu }^{i-}=0 \ ,  \label{de}
\end{align}
actually hold \textit{all along the scalar flow}. In particular, the scalar
fields are \textit{constant} for every value of the radial coordinate $r$.


In this framework, major simplifications take place in the computations. At
the first order, the unique non-trivial variation is given by\footnote{%
Note that from now on the r.h.s. is intended evaluated on the background (%
\ref{de}).}
\begin{equation}
\left. \left( \delta ^{\left( 1\right) }\psi _{A\mu }\right) \right\vert _{%
\mathrm{d.e.}}=\nabla _{\mu }\epsilon _{A}+\varepsilon _{AB}T_{\mu \nu
}^{-}\gamma ^{\nu }\epsilon ^{B}\,,
\end{equation}%
which does not vanish because $\epsilon _{A}$ is an \textit{anti-Killing}
spinor \cite{Aichelburg:1986wv,KL-1,Gentile:2011jt}. Moreover, the subscript
\textquotedblleft $\mathrm{d.e.}$" denotes the evaluation on (\ref{de})%
, throughout. Exploiting the iteration procedure, we then find that at the
next order the bosonic fields are modified as follows:
\begin{align}
\left. \left( \delta ^{\left( 2\right) }e_{\mu }^{a}\right) \right\vert _{%
\mathrm{d.e.}}=& -i\left. \left( \delta ^{\left( 1\right) }\bar{\psi}_{\mu
}^{A}\right) \right. \gamma ^{a}\epsilon _{A}+\mathrm{h.c.}\,,  \notag
\label{A2bg} \\
\left. \left( \delta ^{\left( 2\right) }A_{\mu }^{\Lambda }\right)
\right\vert _{\mathrm{d.e.}}=& 2L^{\Lambda }\left. \left( \delta ^{\left(
1\right) }\bar{\psi}_{\mu }^{A}\right) \right. \epsilon ^{B}\varepsilon
_{AB}+\mathrm{h.c.}\,.
\end{align}%
At the third order, the only non-vanishing variations read
\begin{align}
\left. \left( \delta ^{\left( 3\right) }\psi _{A\mu }\right) \right\vert _{%
\mathrm{d.e.}}=& \left. \left( \delta ^{\left( 2\right) }\nabla _{\mu
}\right) \right. \epsilon _{A}+\left. \left( \delta ^{\left( 2\right)
}T_{\mu \nu }^{-}\right) \right. \gamma ^{\nu }\epsilon ^{B}\varepsilon
_{AB}\,,  \notag \\
\left. \left( \delta ^{\left( 3\right) }\bar{\lambda}^{iA}\right)
\right\vert _{\mathrm{d.e.}}=& -\left. \left( \delta ^{\left( 2\right)
}G_{\mu \nu }^{i-}\right) \right. \bar{\epsilon}_{B}\varepsilon ^{AB}\gamma
^{\mu \nu }\,,
\end{align}%
where
\begin{align}
\left. \left( \delta ^{\left( 2\right) }G_{\mu \nu }^{i-}\right) \right\vert
_{\mathrm{d.e.}}=& -g^{i\bar{j}}\bar{f}_{\bar{j}}^{\Gamma }\mathrm{Im}~%
\mathcal{N}_{\Gamma \Lambda }\left( \delta ^{\left( 2\right) }\tilde{F}_{\mu
\nu }^{\Lambda -}\right) \,,  \notag  \label{F2bg} \\
\left. \left( \delta ^{\left( 2\right) }\tilde{F}_{\mu \nu }^{\Lambda
}\right) \right\vert _{\mathrm{d.e.}}=& \left( \delta ^{\left( 2\right) }%
\mathcal{F}_{\mu \nu }^{\Lambda }\right) +2L^{\Lambda }\left( \delta
^{\left( 1\right) }\bar{\psi}_{\mu }^{A}\right) \left( \delta ^{\left(
1\right) }\psi _{\nu }^{B}\right) \varepsilon _{AB}+\mathrm{h.c.}\,,  \notag
\\
\left. \left( \delta ^{\left( 2\right) }\mathcal{F}_{\mu \nu }^{\Lambda
}\right) \right\vert _{\mathrm{d.e.}}=& \nabla _{\left[ \mu \right. }\left(
\delta ^{\left( 2\right) }A_{\left. \nu \right] }^{\Lambda }\right) \,,
\notag \\
\left. \left( \delta ^{\left( 2\right) }T_{\mu \nu }^{-}\right) \right\vert
_{\mathrm{d.e.}}=& 2i\mathrm{Im}~\mathcal{N}_{\Gamma \Lambda }L^{\Gamma
}\left( \delta ^{\left( 2\right) }\tilde{F}_{\mu \nu }^{\Lambda -}\right) \,,
\notag \\
\left. \left( \delta ^{\left( 2\right) }\mathcal{F}_{\mu \nu }^{\Lambda \pm
}\right) \right\vert _{\mathrm{d.e.}}=& \frac{1}{2}\left( \delta ^{\left(
2\right) }\mathcal{F}_{\mu \nu }^{\Lambda }\right) \pm \frac{i}{4}\left(
\delta ^{\left( 2\right) }\varepsilon _{\mu \nu \rho \sigma }\right)
\mathcal{F}^{\Lambda |\rho \sigma }  \notag \\
\pm & \frac{i}{4}\varepsilon _{\mu \nu \rho \sigma }\left[ g^{\alpha \rho
}g^{\beta \sigma }\left( \delta ^{\left( 2\right) }\mathcal{F}_{\alpha \beta
}^{\Lambda }\right) +2\left( \delta ^{\left( 2\right) }g^{\alpha \rho
}\right) g^{\beta \sigma }\mathcal{F}_{\alpha \beta }^{\Lambda }\right] \,,
\end{align}%
and the same result is obtained for $\tilde{F}_{\mu \nu }^{\Lambda \pm }$.

\section{\label{AM-modified}Modification of the Attractor Mechanism}

By proceeding further with the iteration, one finds that the most relevant
contribution to the variation takes place at the fourth order, at which a
non-vanishing contribution to the variation of the scalar fields is firstly
observed. Thus, the scalar fields get affected by the wigging at the fourth
order in the anti-Killing spinors, even on the simplest background, namely
in the case of double-extremal BPS black hole:
\begin{align}
& \left. \left( \delta ^{\left( 4\right) }A_{\mu }^{\Lambda }\right)
\right\vert _{\mathrm{d.e.}}=2\bar{L}^{\Lambda }\left( \delta ^{\left(
3\right) }\bar{\psi}_{A\mu }\right) \epsilon _{B}\varepsilon
^{AB}+if_{i}^{\Lambda }\left. \left( \delta ^{\left( 3\right) }\bar{\lambda}%
^{iA}\right) \right. \gamma _{\mu }\epsilon ^{B}\varepsilon _{AB}+\mathrm{%
h.c.}\,,  \notag \\
& \left. \left( \delta ^{\left( 4\right) }e_{\mu }^{a}\right) \right\vert _{%
\mathrm{d.e.}}=-i\left. \left( \delta ^{\left( 3\right) }\bar{\psi}_{\mu
}^{A}\right) \right. \gamma ^{a}\epsilon _{A}+\mathrm{h.c.}\,.
\end{align}%
By a long but straightforward algebra, the computation of the fourth-order
variation of the scalar fields can be computed to read:
\begin{equation*}
\left. \left( \delta ^{\left( 4\right) }z^{i}\right) \right\vert _{\mathrm{%
d.e.}}=\left. \left( \delta ^{\left( 4\right) }z_{\nabla }^{i}\right)
\right\vert _{\mathrm{d.e.}}+\left. \left( \delta ^{\left( 4\right)
}z_{T}^{i}\right) \right\vert _{\mathrm{d.e.}}\,,
\end{equation*}%
where we separated two contribution: the one from the spinor covariant
derivative and the one from the graviphoton field-strength
\begin{align}
\left. \left( \delta ^{\left( 4\right) }z_{\nabla }^{i}\right) \right\vert _{%
\mathrm{d.e.}}:=& g^{i\bar{j}}\bar{f}_{\bar{j}}^{\Gamma }\mathrm{Im}~%
\mathcal{N}_{\Gamma \Lambda }\left( \bar{\epsilon}_{C}\gamma ^{\mu \nu
}\epsilon _{D}\right) \varepsilon ^{DC}\left\{ \frac{1}{4}R_{\mu \nu
ab}^{-}L^{\Lambda }\left( \bar{\epsilon}^{A}\gamma ^{ab}\epsilon ^{B}\right)
\varepsilon _{AB}\right.  \notag \\
& -\frac{1}{2}F^{\rho \sigma |\Lambda }\varepsilon _{abcd}\left[ \left(
\nabla _{\mu }\bar{\epsilon}_{A}\gamma ^{a}\epsilon ^{A}\right) e_{\nu
}^{b}e_{\rho }^{c}e_{\sigma }^{d}+e_{\mu }^{a}e_{\nu }^{\nu }\left( \nabla
_{\rho }\bar{\epsilon}_{A}\gamma ^{c}\epsilon ^{A}\right) e_{\sigma }^{d}+%
\mathrm{h.c}\right]  \notag \\
& +\left. F_{\alpha \beta }^{\Lambda }\varepsilon _{\mu \nu \rho }{}^{\beta
}\left( \nabla _{\lambda }\bar{\epsilon}_{A}\gamma ^{c}\epsilon ^{A}+\mathrm{%
h.c}\right) g^{\lambda \left( \rho \right. }e_{c}^{\left. \alpha \right) }%
\phantom{\Big |}\!\right\} \,,  \label{res-1} \\
\left. \left( \delta ^{\left( 4\right) }z_{T}^{i}\right) \right\vert _{%
\mathrm{d.e.}}:=& g^{i\bar{j}}\bar{f}_{\bar{j}}^{\Gamma }\mathrm{Im}~%
\mathcal{N}_{\Gamma \Lambda }\left( \bar{\epsilon}_{C}\gamma ^{\mu \nu
}\epsilon _{D}\varepsilon ^{DC}\right) \left\{ 2L^{\Lambda }\left[ T_{\rho %
\left[ \nu \right. }^{-}\left( \nabla _{\left. \mu \right] }\bar{\epsilon}%
_{A}\gamma ^{\rho }\epsilon ^{A}\right) \right. \right.  \notag \\
& \left. +T_{\rho \left[ \nu \right. }^{-}\left( \nabla _{\left. \mu \right]
}\bar{\epsilon}^{A}\gamma ^{\rho }\epsilon _{A}\right) +\varepsilon
^{AB}T_{\rho \left[ \mu \right. }^{-}T_{\left. \nu \right] \sigma
}^{-}\left( \bar{\epsilon}_{A}\gamma ^{\rho \sigma }\epsilon _{B}\right) %
\right] ^{-}  \notag \\
& -\frac{1}{2}F^{\rho \sigma |\Lambda }\varepsilon _{\rho \nu \omega \sigma
}T_{\lambda \mu }^{-}\left[ \varepsilon _{AB}\left( \bar{\epsilon}^{A}\gamma
^{\lambda \omega }\epsilon ^{B}\right) +\mathrm{h.c.}\right]  \notag \\
& \left. +\frac{1}{2}F_{\rho \sigma }^{\Lambda }\varepsilon _{\mu \nu
}{}^{\sigma \lambda }T_{\lambda \omega }^{-}\ \left[ \varepsilon _{AB}\left(
\bar{\epsilon}^{A}\gamma ^{\rho \omega }\epsilon ^{B}\right) +\mathrm{h.c.}%
\right] \right\} \ ,  \label{res-2}
\end{align}
where we defined
\begin{equation}
R_{\mu \nu ab}^{-}:=\frac{1}{2}\left( R_{\mu \nu ab}-\frac{i}{2}\varepsilon
_{\mu \nu }{}^{\rho \sigma }R_{\rho \sigma ab}\right) \ .
\end{equation}%
and 

Since%
\begin{equation}
\left. \left( \delta ^{\left( 1\right) }z^{i}\right) \right\vert _{\mathrm{%
d.e.}}=\left. \left( \delta ^{\left( 2\right) }z^{i}\right) \right\vert _{%
\mathrm{d.e.}}=\left. \left( \delta ^{\left( 3\right) }z^{i}\right)
\right\vert _{\mathrm{d.e.}}=0,
\end{equation}%
it thus follows that the complete fermionic wig of the $n$ complex scalar
fields $z^{i}$ in the background of a double-extremal {1}/{2}-BPS black hole
in $\mathcal{N}=2$, $D=4$ \textit{minimally coupled} supergravity reads (in
absence of gauging and hypermultiplets):
\begin{equation}
\left. z_{WIG}^{i}\right\vert _{\mathrm{d.e.}}:=\left.
z_{(0)}^{i}\right\vert _{\mathrm{d.e.}}+\frac{1}{4!}\left. \left( \delta
^{\left( 4\right) }z^{i}\right) \right\vert _{\mathrm{d.e.}}\neq \left.
z_{(0)}^{i}\right\vert _{\mathrm{d.e.}},  \label{AM-Wig}
\end{equation}%
where $\left. z_{(0)}^{i}\right\vert _{\mathrm{d.e.}}$ denotes the \textit{%
\textquotedblleft unwigged"}, near-horizon value of the scalar fields;
according to the {attractor mechanism} \cite{AM-Refs}, the latter depends
only on the electric and magnetic charges of the black hole (for a detailed
treatment, see \cite{Gnecchi-1}, and references therein). \vspace{0.2cm}

\section{\label{Axion-Dilaton}Axion-Dilaton Model}

As an illustrative example, we analyze the simplest case within \textit{%
minimally coupled} $\mathcal{N}=2$ supergravity, namely the $\overline{%
\mathbb{CP}}^{1}$ model, with only one vector multiplet (containing one
complex scalar field $z$) coupled to the gravity multiplet.

In this case, we find convenient to consider the symplectic frame specified
by the holomorphic prepotential%
\begin{equation}
F:=-iX^{0}X^{1}\ ,  \label{F-ad}
\end{equation}%
which
arises out by suitably truncating the $\mathcal{N}=4$ \textquotedblleft
pure" theory (see \textit{e.g.} the discussion in \cite{Hayakawa-1,FMO-MS-1}%
), and it determines the following K\"{a}hler potential (\textit{cfr. e.g.}
\cite{Kallosh:1996tf}):
\begin{equation}
K=-\ln \left[ 2\left( z+\bar{z}\right) \right] \ ,  \label{K-ad}
\end{equation}%
from which the metric function is derived:
\begin{equation}
g_{1\bar{1}}=\left( g^{1\bar{1}}\right) ^{-1}=\frac{1}{\left( z+\bar{z}%
\right) ^{2}}\ .
\end{equation}

In special coordinates, after a K\"{a}hler gauge-fixing ($\Lambda =0,1$
throughout the present Section):
\begin{equation}
X^{\Lambda }=\left( 1,z\right) \ ,
\end{equation}%
and then one can derive the covariantly holomorphic symplectic sections of
special geometry:
\begin{align}
& L^{\Lambda }:=e^{\frac{K}{2}}X^{\Lambda }=\frac{1}{\sqrt{2\left( z+\bar{z}%
\right) }}\left( 1,z\right) \,, \\
& M_{\Lambda }:=\mathcal{N}_{\Lambda \Sigma }L^{\Sigma }=-i\frac{1}{\sqrt{%
2\left( z+\bar{z}\right) }}\left( z,1\right) \ ,
\end{align}%
and their K\"{a}hler-covariant derivatives
\begin{equation}
f^{\Lambda }:=\left( \partial _{z}+\frac{1}{2}\partial _{z}K\right)
L^{\Lambda }=\frac{1}{\sqrt{2}\left( z+\bar{z}\right) ^{3/2}}\left( -1,\bar{z%
}\right)
\end{equation}%
(note the suppression of the $i$-index in $f_{i}^{\Lambda }$, due to the
presence of only one scalar field).

In a symplectic frame defined by a prepotential $F$, the symmetric complex
kinetic matrix of vector fields is defined as (see for instance \cite%
{Ceresole-1,Ceresole-2}, and Refs. therein)
\begin{align}
\mathcal{N}_{\Lambda \Sigma }& :=\bar{F}_{\Lambda \Sigma }-2i\bar{T}%
_{\Lambda }\bar{T}_{\Sigma }\left( L^{\Gamma }\mathrm{Im}F_{\Gamma \Xi
}L^{\Xi }\right) , \\
F_{\Lambda \Sigma }& :=\frac{\partial ^{2}F}{\partial X^{\Lambda }\partial
X^{\Sigma }}\ , \\
T_{\Lambda }& :=-i\frac{\mathrm{Im}F_{\Lambda \Xi }\bar{L}^{\Xi }}{\bar{L}%
^{\Gamma }\mathrm{Im}F_{\Gamma \Sigma }\bar{L}^{\Sigma }}.
\end{align}

In the case under consideration, the $2\times 2$ kinetic vector matrix reads
\begin{equation}
\mathcal{N}_{\Lambda \Sigma }=-i\ \mathrm{diag}\left( z,\frac{1}{z}\right) ,
\end{equation}%
thus yielding%
\begin{align}
\mathrm{Im}~\mathcal{N}_{\Lambda \Sigma } &= -\frac{z+\bar{z}}{2}\mathrm{diag%
}\left( 1,\frac{1}{|z|^{2}}\right) , \\
\mathrm{Re}~\mathcal{N}_{\Lambda \Sigma } &= \frac{z-\bar{z}}{2 i}\mathrm{%
diag}\left( 1,-\frac{1}{|z|^{2}}\right) .
\end{align}

\subsection{\label{D-Extr-BH}Double--Extremal Black Hole}

We are now going to derive the explicit values for the various fields in our
configuration. We will be dealing with an asymptotically flat, static,
spherically symmetric, dyonic $1/2-$BPS \textit{double--extremal }black
hole, with $z$ constant for every value of the radial coordinate $r$.
Following the conventions of \cite{Kallosh:1996tf}, we consider a dyonic
black hole metric\footnote{%
This metric is of Papetrou-Majumdar form, thus the radius of the event
horizon is located at $r=r_{H}=0$.}
\begin{equation}
ds^{2}=\left( 1+\frac{M}{r}\right) ^{-2}dt^{2}-\left( 1+\frac{M}{r}\right)
^{2}\left( dr^{2}+r^{2}d\theta ^{2}+r^{2}\sin ^{2}\theta d\phi ^{2}\right)
\,,  \label{ds-ad}
\end{equation}%
with gauge field strengths given by
\begin{equation}
\mathcal{F}^{\Lambda }=\left( 1+\frac{M}{r}\right) ^{-2}\frac{2Q^{\Lambda }}{%
r^{2}}dt\wedge dr-2P^{\Lambda }\sin \theta d\theta \wedge d\phi \ ,
\label{gauge-field-ad}
\end{equation}%
where $Q^{\Lambda }$ and $P^{\Lambda }$ are symplectic, $z$-dependent, real
quantities, defined by \cite{Kallosh:1996tf}%
\begin{equation}
{\binom{P^{\Lambda }}{Q^{\Lambda }}}=\frac{1}{2}{\binom{p^{\Lambda }}{\left(
\mathrm{Im}~\mathcal{N}^{-1}\mathrm{Re}~\mathcal{N}\ p\right) ^{\Lambda
}-\left( \mathrm{Im}~\mathcal{N}^{-1}\ q\right) ^{\Lambda }}}\ .
\label{rell}
\end{equation}

As showed in the third Ref. of \cite{AM-Refs}, in order to have a
supersymmetric attractor solution one must require that $G_{\mu \nu }^{i-}=0$
on the horizon; such a requirement constrains the scalar $z$ to be a
function only of the electric ($q_{\Lambda }$) and magnetic ($p^{\Lambda }$)
charges of the black hole. Starting from (\ref{K-ad}), it turns out that the
value of the constant scalar is fixed to be
\begin{equation}
\left. z^{(0)}\right\vert _{\mathrm{d.e.}}=\frac{q_{0}-ip^{1}}{q_{1}-ip^{0}}%
=\,\left. \frac{Q^{1}-iP^{1}}{Q^{0}-iP^{0}}\right\vert _{\mathrm{d.e.}},
\label{zorder0}
\end{equation}%
where in the last step the inverse of (\ref{rell}), namely \cite%
{Kallosh:1996tf}%
\begin{equation}
{\binom{p^{\Lambda }}{q_{\Lambda }}}={\binom{2P^{\Lambda }}{2\mathrm{Re}%
\mathcal{N}_{\Lambda \Sigma }P^{\Sigma }-2\mathrm{Im}~\mathcal{N}_{\Lambda
\Sigma }Q^{\Sigma }},}\   \label{rell-2}
\end{equation}%
has been exploited. Note that $\left. z^{(0)}\right\vert _{\mathrm{d.e.}}$%
given by (\ref{zorder0}) expresses the value of the scalar field $z$ at the
\textit{zeroth} order.

By setting $z=\left. z^{(0)}\right\vert _{\mathrm{d.e.}}$, one gets $G_{\mu
\nu }^{i-}=0$, and the BPS bound is saturated \cite{Kallosh:1996tf}:
\begin{eqnarray}
M^{2} &=&-2\left[ \mathrm{Im}~\mathcal{N}_{\Gamma \Lambda }\left( Q^{\Gamma
}Q^{\Lambda }+P^{\Gamma }P^{\Lambda }\right) \right] _{\mathrm{d.e.}%
}=\left\vert Z\right\vert ^{2}\,_{\mathrm{d.e.}}  \notag \\
&=&q_{0}q_{1}+p^{0}p^{1}=\frac{A_{H(0)}}{4}=\frac{S_{BH(0)}}{\pi }\ ,
\label{BPS-bound}
\end{eqnarray}%
where $Z$ is the $\mathcal{N}=2$ central charge function:
\begin{equation}
Z:=L^{\Lambda }q_{\Lambda }-M_{\Lambda }p^{\Lambda }\ ,
\end{equation}%
and $A_{H(0)}$ and $S_{BH(0)}$ respectively denote the horizon area and the
Bekenstein-Hawking entropy of the black hole at the \textit{zeroth order}
(recall the comment at the end of Sec.~\ref{Intro}).

In order for the scalar to be fixed at order $n$ (in particular $n=4$), one
has to require the vanishing of the supersymmetry variation $\delta ^{\left(
n-2\right) }G_{\mu \nu }^{i-}$. As shown below, due to the presence of a
gauge field variation, this is true only up to $n=3$.

\subsection{\label{4th-Order}Fourth Order Scalar Variation}

We now proceed to computing the fourth order variation of the scalar field $%
z $ in the double-extremal BPS axion-dilaton background specified by (\ref%
{ds-ad}), (\ref{gauge-field-ad}) and (\ref{zorder0}), as described in Sec.~%
\ref{D-Extr-BH}.

We start and recall the Minkowski-Killing spinors in spherical coordinates
:
\begin{align}
& \epsilon _{A}=\left[ \cos \frac{\theta }{2}\left( \sin \frac{\phi }{2}%
\mathrm{1\kern-.9mml}_{4}+\cos \frac{\phi }{2}\gamma _{23}\right) +\sin
\frac{\theta }{2}\left( \cos \frac{\phi }{2}\gamma _{13}-\sin \frac{\phi }{2}%
\gamma _{12}\right) \right] \zeta _{A}\,, \\
& \epsilon ^{A}=\left[ \cos \frac{\theta }{2}\left( \sin \frac{\phi }{2}%
\mathrm{1\kern-.9mml}_{4}+\cos \frac{\phi }{2}\gamma _{23}\right) +\sin
\frac{\theta }{2}\left( \cos \frac{\phi }{2}\gamma _{13}-\sin \frac{\phi }{2}%
\gamma _{12}\right) \right] \zeta ^{A}\,,
\end{align}%
where $\zeta _{1}=\frac{\left( 1+\gamma _{5}\right) }{2}\mathbf{1}$, $\zeta
_{2}=\frac{\left( 1+\gamma _{5}\right) }{2}\mathbf{2}$, $\zeta ^{1}=\frac{%
\left( 1-\gamma _{5}\right) }{2}\mathbf{1}$, $\zeta ^{2}=\frac{\left(
1-\gamma _{5}\right) }{2}\mathbf{2}$ and $\mathbf{1},\mathbf{2}$ are
Majorana spinors defined as
\begin{equation}
\mathbf{1}=\left\{ a_{1},a_{2},-a_{2}^{\ast },-a_{1}^{\ast }\right\}
\,,\qquad \mathbf{2}=\left\{ b_{1},b_{2},-b_{2}^{\ast },-b_{1}^{\ast
}\right\} \,,
\end{equation}%
with $a,b$ denoting \textit{constant} complex Grassmannian numbers.

As mentioned above, the non-vanishing variation for the scalar field $z$ is
induced by the correction that $G_{\mu \nu }^{i-}$ \textit{does} \textit{%
acquire} at the second order. In fact, one achieves the following result:
\begin{equation}
\left. \left( \delta ^{\left( 4\right) }z^{i}\right) \right\vert _{\mathrm{%
d.e.}}=\left. \left( \delta ^{\left( 3\right) }\bar{\lambda ^{iA}}\right)
\right\vert _{\mathrm{d.e.}}\epsilon _{A}=-\left. \left( \delta ^{\left(
2\right) }G_{\mu \nu }^{i-}\right) \gamma ^{\mu \nu }\right\vert _{\mathrm{%
d.e.}}\bar{\epsilon}_{B}\epsilon _{A}\varepsilon ^{AB}\,,
\end{equation}%
(\textit{cfr.} Sec.~\ref{Wigging} for the explicit variation of the fields);
we should note that we exploited the special geometry identity (see \textit{%
e.g.} \cite{N=2-Big})
\begin{equation}
\mathrm{Im}~\mathcal{N}_{\Lambda \Sigma }\bar{f}_{\bar{j}}^{\Lambda }\bar{L}%
^{\Sigma }=0\,.
\end{equation}

\subsection{\label{Final-Result}Final Result}

By recalling the results (\ref{res-1}) and (\ref{res-2}), the value of the
scalar field at the fourth order (\ref{zorder0}) yields
\begin{equation}
\left. \left( \delta ^{\left( 4\right) }z\right) \right\vert _{\mathrm{d.e.}%
}=\left. \left( \delta ^{\left( 4\right) }z_{T}\right) \right\vert _{\mathrm{%
d.e.}}\ .  \label{zat4final}
\end{equation}

It should be stressed that, upon acting with all vacuum super-isometries as
supersymmetry parameters, $\left. \left( \delta ^{\left( 4\right) }z\right)
\right\vert _{\mathrm{d.e.}}$ acquires a dependence also on the \textit{%
unbroken} super-isometries. This redundance can be eliminated by a gauge
choice on the gravitino field, in order to work with a \textquotedblleft
pure\textquotedblright\ anti--Killing spinor with $4$ (complex) degrees of
freedom. In order to highlight their contribution, we redefine the \textit{%
costant} Minkowski-Killing spinor zero modes as follows:
\begin{align}
& A:=a_{1}+ib_{1}\,,\qquad B:=b_{2}-ia_{2}\,,  \notag \\
& C:=a_{1}^{\ast }+ib_{1}^{\ast }\,,\qquad D:=b_{2}^{\ast }-ia_{2}^{\ast }\ .
\end{align}%
Such a redefinition allow us to work only with $A,B,C$ and $D$, since these
are the only generators for the black hole wig itself (their complex
conjugates are the zero modes for the black hole Killing spinors). Using
these variables, we finally achieve the result
\begin{eqnarray}
\left. \left( \delta ^{\left( 4\right) }z\right) \right\vert _{\mathrm{d.e.}%
} &=&\frac{M^{4}}{4\left( M+r\right) ^{4}}\left[ \frac{P^{0}Q^{1}-P^{1}Q^{0}%
}{\left( P^{0}+iQ^{0}\right) ^{2}\left( Q^{0}+iP^{0}\right) \left(
P^{1}-iQ^{1}\right) }\right] _{\mathrm{d.e.}}\mathbf{\mathcal{Q}}\sin
^{2}\phi \sin ^{2}\theta \   \label{pre-result} \\
&=&\frac{M^{4}}{\left( M+r\right) ^{4}}\frac{p^{0}q_{0}-p^{1}q_{1}}{\left(
p^{0}+iq_{1}\right) ^{2}\left( p^{0}-iq_{1}\right) \left(
q_{0}+ip^{1}\right) }\mathbf{\mathcal{Q}}\sin ^{2}\phi \sin ^{2}\theta ,
\label{result}
\end{eqnarray}%
within the constraint $q_{0}q_{1}+p^{0}p^{1}>0$ imposed by the saturation of
the BPS bound (\ref{BPS-bound}). Note that we have introduced the \textit{%
\textquotedblleft quadrilinear\textquotedblright }\ $\mathbf{\mathcal{Q}}%
:=ABCD$, and Eqs. (\ref{rell}) and (\ref{zorder0}) have been used. Also, for
$M=0$ the result (\ref{result}) vanishes, as expected.

By evaluating the expression (\ref{result}) on the event horizon $r=r_{H}=0$
of the bosonic solution (\ref{ds-ad}) (denoted by the subscript
\textquotedblleft $\mathrm{d.e.h.}$"; recall the comment at the end of Sec.~%
\ref{Intro}), one obtains
\begin{equation}
\left. \left( \delta ^{\left( 4\right) }z\right) \right\vert _{\mathrm{d.e.h.%
}}=\frac{p^{0}q_{0}-p^{1}q_{1}}{\left( p^{0}+iq_{1}\right) ^{2}\left(
p^{0}-iq_{1}\right) \left( q_{0}+ip^{1}\right) }\mathbf{\mathcal{Q}}\sin
^{2}\phi \sin ^{2}\theta \ ,  \label{result-2}
\end{equation}

As resulting from~(\ref{result-2}) and (\ref{AM-Wig}), in the \textit{%
near-horizon} background of a double--extremal BPS axion-dilaton black hole,
upon performing (the near-horizon limit of) a finite supersymmetry
transformation, the axion-dilaton $z$ \textit{is not constant any more}, but
acquires a \textit{dependence on the angles }$\phi $ and $\theta $.

Nevertheless, for $M\neq 0$, one can single out \textit{at least} three
peculiar charge configurations in which $z$ \textit{does} remain fixed, and
given by (\ref{zorder0}), \textit{i.e.} in which\footnote{%
Note that (\ref{result}) and its horizon limit (\ref{result-2}) only differ
by the $r$-dependent pre-factor $M^{4}/\left( M+r\right) ^{4}$. Furthermore,
$\left. z^{(0)}\right\vert _{\mathrm{d.e.h.}}=\left. z^{(0)}\right\vert _{%
\mathrm{d.e.}}$, because we are considering a \textit{double-extremal}
bosonic solution (\ref{ds-ad}).} $\left. \left( \delta ^{\left( 4\right)
}z\right) \right\vert _{\mathrm{d.e.}}=0=\left. \left( \delta ^{\left(
4\right) }z\right) \right\vert _{\mathrm{d.e.h.}}$:%
\begin{equation}
\left.
\begin{array}{l}
\mathbf{I.~}\ p^{0}=p^{1}=0\Longrightarrow \left. z^{(0)}\right\vert _{%
\mathrm{d.e.}}=q_{0}/q_{1}; \\
\phantom{ciao} \\
\mathbf{II.~}\ q_{0}=q_{1}=0\Longrightarrow \left. z^{(0)}\right\vert _{%
\mathrm{d.e.}}=p^{1}/p^{0}\ ; \\
\phantom{ciao} \\
\mathbf{III.~}\ p^{1}/p^{0}=q_{1}/q_{0}.%
\end{array}%
\right\} \Rightarrow \left. z_{WIG}\right\vert _{\mathrm{d.e.}}=\left.
z_{WIG}\right\vert _{\mathrm{d.e.h.}}=\left. z^{(0)}\right\vert _{\mathrm{%
d.e.h.}}=\left. z^{(0)}\right\vert _{\mathrm{d.e.}}
\end{equation}

\section{\label{Conclusion}Conclusions}

Eq.~(\ref{AM-Wig}), with $\left. \left( \delta ^{\left( 4\right)
}z^{i}\right) \right\vert _{\mathrm{d.e.}}$ given by the results (\ref%
{pre-result})-(\ref{result}) and (\ref{res-2}), expresses how the value of
the axion-dilaton gets \textit{modified by the fermionic wig along the
radial flow} in the background of a bosonic BPS double extremal black hole
of $\mathcal{N}=2$ supergravity.

In particular, its near-horizon limit, in which the expressions (\ref%
{pre-result})-(\ref{result}) are replaced by (\ref{result-2}), yields that
the \textit{attractor mechanism gets modified by the fermionic wig}. It is
therefore the first evidence - in the simplest case provided by the (double
extremal) axion-dilaton black hole, of what we dub the \textit{%
\textquotedblleft fermionic-wigged" attractor mechanism}: the \textit{%
fermionic-wigged} value, depending on the \textquotedblleft quadrilinear" $%
\mathcal{Q}$ as well as on the angles $\phi $ and $\theta $, of the scalar
fields in the near-horizon geometry of the double-extremal $1/2$-BPS black
hole is \textit{different} from the corresponding, purely charge--dependent,
horizon attractor value at the \textit{zeroth} order.

We would like to stress once again that we adopted the approximation of
computing the fermionic wig by performing a perturbation of the unwigged,
purely bosonic (double) extremal BPS extremal black hole solution; thus,
within this approximation, we consider quantities like the radius of the
event horizon unchanged.

We leave to further future work \cite{to-appear-1} the complete analysis of
the fully-backreacted wigged black hole solution, also including the study
of its thermodynamical properties, and the computation of its
Bekenstein-Hawking entropy; this may be done also in the non-supersymmetric
(non-BPS) case.

Our analysis may also be applied to higher dimensions, as well as to
extended supergravities.

\section*{Acknowledgments}

We would like to thank S. Ferrara, M. Porrati and L. Sommovigo for
enlightening discussions. A. Marrani would like to thank the DISIT for kind
hospitality.

The work of PAG is partially supported by the MIUR-PRIN contract
2009-KHZKRX. The work of A. Marrani is supported in part by the FWO -
Vlaanderen, Project No. G.0651.11, and in part by the Interuniversity
Attraction Poles Programme initiated by the Belgian Science Policy (P7/37).
The work of A. Mezzalira is partially supported by IISN - Belgium
(conventions 4.4511.06 and 4.4514.08), by the \textquotedblleft Communaut%
\'{e} Fran\c{c}aise de Belgique" through the ARC program and by the ERC
through the \textquotedblleft SyDuGraM" Advanced Grant.

\end{document}